\documentclass[reprint,amsmath,amssymb,aps,prl,groupedaddress,superscriptaddress]{revtex4-1}

\usepackage{graphicx}
\usepackage{dcolumn}
\usepackage{bm}

\newcommand{\be}{\begin{equation}}
\newcommand{\ee}{\end{equation}}
\def \ba {\begin{aligned}}
\def \ea {\end{aligned}}

\newcommand{\nn}{\nonumber}
\def \th {\theta}
\def \L {\Lambda}

\begin{document}

\title{A new method for exact results on Quasinormal Modes of Black Holes}

\author{Davide Fioravanti}
\email[]{fioravanti at bo.infn.it}
\affiliation{Sezione INFN di Bologna and Dipartimento di Fisica e Astronomia, Universit\`a di Bologna \\
Via Irnerio 46, 40126 Bologna, Italy}

\author{Daniele Gregori}
\email[]{daniele.gregori6 at unibo.it}
\altaffiliation{also visiting NORDITA.}
\affiliation{Sezione INFN di Bologna and Dipartimento di Fisica e Astronomia, Universit\`a di Bologna \\
Via Irnerio 46, 40126 Bologna, Italy}


\begin{abstract}
\noindent We develop a new method for writing simple exact equations characterizing gravity solutions among which black holes and in particular the quasinormal modes. More precisely,  we derive the full system of functional and Thermodynamic Bethe Ansatz non linear integral equations of quantum integrability. In particular, we prove that the Quasinormal Modes verify different equivalent exact quantization conditions and identify them with Bethe roots. We numerically solve the integral equation and compare the results with other methods. Eventually, we can definitely certify its simplicity, accuracy and effectiveness. Furthermore, this method connects different unexpected fields and paves the way for innovative ways of investigations in gravity and gauge theories. 
\end{abstract}


\maketitle

\section{I. Introduction, plan and goals}\label{intro-plan}

Thanks to the extraordinary discovery of gravitational waves in 2015~\cite{LIGOScientific:2016}, a new window for our understanding of the Universe has opened. In fact, it has become possible to make progress in also fundamental physics, testing General Relativity (GR) in extreme regimes and in particular to discriminate between GR Black Holes (BHs) and Exotic Compact Objects (ECOs) or Fuzzballs 
appearing in Modified Theories of Gravity or String Theory. This is possible importantly by analysing the Quasinormal Modes (QNMs) that characterize the linear perturbations of the metric and the fields, for instance in the ringdown phase of BHs merging~\cite{Cardoso:2017cqb,BianchiConsoliGrilloMorales:2021,Mayerson:2020}. Computing QNMs numerically has been until now typically quite laborious and this has been also due to the difficulties in developing exact analytic characterizations of QNMs. In this direction, a significant improvement has been realized very recently as QNMs have been identified by exact quantization conditions on {\it dual} periods of some $\mathcal{N}=2$ supersymmetric gauge, {\it i.e.} deformed Seiberg-Witten (SW), theories~\cite{AminovGrassiHatsuda:2020,SeibergWitten:1994b,NekrasovShatashvili:2009}. On the latter we have in fact some exact control and then this surprising SW-QNM correspondence -- so dubbed in \cite{BianchiConsoliGrilloMorales:2021b} -- has already allowed to find many new theoretical and computational results for BHs and other spacetime geometries ({\it cf.} for instance~\cite{CasalsCosta:2021,Hatsuda:2020Teukolsky,BonelliIossaLichtigTanzini:2021}). An explanation of this correspondence has been constructed in a rather general case~\cite{BonelliIossaLichtigTanzini:2021} by exploiting another correspondence between $\mathcal{N}=2$ gauge theory and Conformal Field Theory~\cite{AldayGaiottoTachikawa:2010}. However, we are going to show that it is possible to explain the SW-QNM correspondence by analysing closely the Ordinary Differential Equations (ODEs) describing the perturbations in gravitational physics. We are able to do this on the basis of our previous works~\cite{FioravantiGregori:2019,FioravantiPoghossian:2019}, where we have connected the $\mathcal{N}=2$ gauge theories to quantum integrable theories, in particular the gauge periods to the Baxter's $Q$ and $T$ functions. To this aim we have started from the ODEs characterizing the periods and developed further the elegant ODE/IM correspondence between ODEs and Integrable Models (IM)~\cite{DoreyTateo1998,BazhanovLukyanovZamolodchikov2001,DoreyTateo1999,FioravantiRossi:2021}.

Here we prove that QNMs are nothing but the zeros of the $Q$ function (Bethe roots) and then find an entirely new set of functional and integral (non-linear) equations for them (in particular quantization conditions). In specific, we derive a new quantization condition involving the solution of a celebrated non-linear integral equation, the Thermodynamic Bethe Ansatz (TBA) ones. As a result, this way turns out to be a very simple and powerful way to compute QNMs. In the light of these findings, we understand and develop here the lines of generality in our construction~\cite{FioravantiGregori:2019,FioravantiPoghossian:2019} to put it into correspondence with gravitational physics.

In this letter, after introducing QNMs in section II, in section III we explain the Integrability/Gauge/Gravity correspondence in the simplest case of the D3 brane spacetime, corresponding to our previous treatment of $SU(2)$ $N_f=0$ gauge theory via the self-dual Liouville theory~\cite{FioravantiGregori:2019}. To show the generality of the Integrability/Gravity correspondence, in section IV we develop the same theory for the spacetime given by the intersection of D3 branes, a generalization of extremal Reissner-Nordstr\"om (RN) BHs, corresponding to $SU(2)$ $N_f=2$ gauge theory~\cite{Ikeda:2021,BianchiConsoliGrilloMorales:2021b} and a new IM (which reproduces known IMs in some limits, {\it cf.} below). In section V we highlight the reasons for applicability of our method and conclude that different features of many Integrability/Gauge/Gravity cases can be treated in future perspective.

\section{II. Brief overview on QNMs}\label{defQNMs}

We recall the definition of quasinormal modes following~\cite{Nollert:1999}. A linear perturbation of a BH is a solution $\Phi(t,x)$ of some linear PDE derived from the equations for the fields and metric. Its Laplace transform $\hat{\Psi}(s,x) $, with $\Re s > 0$, satisfies some ODE with non-homogeneous term $\mathcal{I}(s,x)$, combination of $\Phi$ and its time derivative at initial time. The corresponding homogeneous equation is exactly the ODE we are going to study in the next sections, which we write here for general potential $U(x)$
\be  \label{homogeneousODE}
\left \{-\frac{\partial ^2}{\partial x^2}+U(x)+s^2\right \} \Psi(s,x) =0 \,.
\ee 
As usual, $x$ is the so called {\it tortoise} coordinate, such that the BH horizon is put at $x \to -\infty$ and spacetime infinity at $x \to +\infty$. 
For decaying potentials, the bounded solutions behave at $x \to \pm \infty $ respectively as
\begin{align}  \label{FundRegSolGen}
\Psi_\pm(s,x) &\sim e^{\mp s x}\, \,.
\end{align} 
Via those known as Green functions, we can write the solution of the non-homogenous equation $\hat{\Psi}(s,x) $ and then take its inverse Laplace transform to obtain the original perturbation as 
\small\begin{align}
&\Phi(t,x)= \\
&\sum_n e^{s_n t} \text{Res}\left( \frac{1}{W(s)}\right)\Biggr|_{s_n}\int_{-\infty}^\infty \Psi_-(s_n,x_<) \Psi_+(s_n, x_>) \mathcal{I}(s_n,x') \, d x' \nn
 \label{QNMsum}
\end{align}
\normalsize
where \small$x_<= \text{min}(x',x) \,, x_<= \text{max}(x',x)\,$\normalsize.
The crucial point for us is that the perturbation is a sum over the residues of the inverse wronskian of the regular solutions~\eqref{FundRegSolGen}: 
\be  \label{QNMcond}
W(s_n)=W[\Psi_+,\Psi_-] =0 \,.
\ee 
Besides, condition~\eqref{QNMcond} means that at these special points the two solutions~\eqref{FundRegSolGen} (in general independent) become linearly dependent. By setting $s = i \omega$ we recover the usual intuitive definition of QNMs as the frequencies of plane wave solutions both incoming at the horizon and outgoing at infinity. However, as well explained in~\cite{Nollert:1999}, this last definition is not mathematically rigorous, since it would lead to diverging boundary conditions. Instead, the QNMs $\omega_n$ have an imaginary part $\Im \omega_n <0$, so that the perturbation they describe is damped to zero as $t \to \infty$.

\section{III. An example: the D3 brane}
\label{D3brane}

The D3 brane is described by the line element
\small
\be 
d s^2 = H(r)^{-\frac{1}{2}} (- dt^2 + d \mathbf{x}^2)+ H(r)^{\frac{1}{2}}(d r^2+ r^2 d \Omega_5^2)\,,
\ee
\normalsize
where $\mathbf{x}$ are the longitudinal coordinates, \small$H(r) = 1 + L^4/r^4$\normalsize and $d \Omega_5^2$ denotes the metric of the transverse round $S^5$-sphere~\cite{BianchiConsoliGrilloMorales:2021}. The ODE which describes the scalar field perturbation of the D3 brane is~\cite{BianchiConsoliGrilloMorales:2021,GubserHashimoto:1998}
\small
\be \label{ODED3r} 
\frac{d^2 \phi}{d r^2} + \left[\omega^2 \left(1+ \frac{L^4}{r^4}\right)-\frac{(l+2)^2-\frac{1}{4}}{r^2} \right]\phi=0\,.
\ee 
\normalsize
Upon the change of 
variables
\small
\be \label{par0}
r = L e^{\frac{y}{2}} \,\quad \omega L =-2 i e^{\theta} \,\quad P=\frac{1}{2}( l+2) \,,
\ee
\normalsize
the equation reduces to the generalized  Mathieu equation
\be \label{ODEsdLiouv}
-\frac{d^2}{d y^2}\psi + \left[e^{2\theta} (e^{y}+e^{-y}) + P^2  \right]\psi = 0\,.
\ee 
It corresponds to the conformal self-dual Liouville theory with momentum $P$ and rapidity $\th$~\cite{FioravantiGregori:2019,ZamolodchikovMemorial}. Instead, to obtain the $\mathcal{N}=2$ $SU(2)$ pure ($N_f=0$) gauge theory with Omega background in the Nekrasov Shatashvili limit $\epsilon_2 \to 0$, $\epsilon_1 = \hslash \neq 0$, we need to use the change of parameters
$\omega L =-2 i  \frac{\L_0}{\hslash} \,, \quad   \frac{1}{8}(l+2)^2=\frac{u}{\hslash^2}$, where $u$ is the Coulomb branch modulus and $\L_0$ the instanton coupling~\cite{FioravantiGregori:2019}. Here we will just state the necessary results of the ODE/IM correspondence procedure for the Liouville Field Theory and refer to~\cite{FioravantiGregori:2019} for details and the proper 2D statistical interpetation. In fact, we want to explain ODE/IM correspondence for a more general case in section III. In general, the starting point of the ODE/IM method is the definition of a $Q$ function as the wronskian of two regular solutions $\psi_{+,0},\psi_{-,0}$ at $y \to \pm \infty$, respectively:
\be
Q(\th,P) = W[\psi_{+,0},\psi_{-,0}](\th,P)\,.
\ee 
Of course, $\psi_{+,0}(y) = \psi_{-,0}(-y)$ and {\it cf.} next section for physical interpretation. Then we derived the fundamental $QQ$ functional relation
\begin{equation} 
 Q(\theta + i \pi/2) Q(\theta -i \pi/2)= 1+ Q(\theta)^2  \, , \label{QQ0}
\end{equation}
from which we derived all the theory. Here and in the following we can omit the dependence on $P$ as it stays fixed. Crucially, the QNMs condition~\eqref{QNMcond} translates into
\be  \label{QNMBethe}
Q(\th_n) =0 \,,
\ee 
namely the zeros of the Baxter's $Q$ function which are the Bethe roots~\cite{Bazhanov:1996dr}. Now, we prove that condition~\eqref{QNMBethe} is equivalent to the quantization condition of the dual gauge period
\small
\be  \label{QuantAD0}
A_D(a,\L_{0,n},\hslash) = i\pi  n \,,\qquad n \in \mathbb{Z}\,.
\ee 
\normalsize
as conjectured by~\cite{AminovGrassiHatsuda:2020}. For it was already proposed in~\cite{GrassiGuMarino} on a numerical basis this relevant relation 
\small
\be  \label{Q0GGM}
Q(a, \L_0,\hslash) =i \frac{\sinh A_D(a,\L_0,\hslash)}{\sinh  \frac{2\pi i}{\hslash} a} \,,
\ee 
\normalsize 
involving the period $a$ and the dual one $A_D=\partial{\mathcal{F}}/{\partial a}$. Actually, it can be proven within ODE/IM or checked numerically by easily computing the l.h.s. with TBA below (\ref{TBA0}). Instead, the computation of $A_D$ in (\ref{QuantAD0},\ref{Q0GGM}) relies on the expansion of the prepotential $\mathcal{F}$ in powers (number of instantons) of $\L_0^4$~\cite{Nekrasov:2002,NekrasovOkounkov:2003}: the period $a$ is related to the moduli parameter $u$ (or $P$) through the Matone's relation~\cite{Matone:1995,FlumeFucitoMoralesPoghossian:2004}. In this respect, only the first instanton contributions are easily accessible and summing them up (naively) is accurate as long as $|\L_0|/\hslash \ll 1$. Thus, this make hard to access QNMs values $|\L_{0,n}|/\hslash \gg 1$.

On the contrary, we found in general this procedure to an exact result, which resums all instantons. Let us define the $Y(\theta,P) =Q^2(\theta,P)$ function and derive from~\eqref{QQ0} the $Y$-system
\begin{equation} 
Y(\theta + i \pi /2)Y(\theta - i \pi /2) = \Bigl (1+Y(\theta )\Bigr)^2  \label{Ysystem0} \, .
\end{equation} 
Eventually, we solve it explicitly ({\it up to quadratures}) via a Thermodynamic Bethe Ansatz (TBA) integral equation for the pseudoenergy $\varepsilon(\theta) = - \ln Y(\theta)$:
\small
\begin{align} 
\varepsilon(\theta)   &=  \frac{16\sqrt{\pi^3}}{\Gamma(\frac{1}{4})^2} e^{\theta} - 2 \int_{-\infty}^\infty\frac{ \ln \left [ 1 +  \exp \{ - \varepsilon(\theta') \} \right ] }{\cosh (\theta-\theta' ) } \frac{d \theta'}{ 2 \pi} \, .
\label{TBA0}
\end{align} 
\normalsize
In this $P$ does not appear explicitly, but fixes the solution by its asymptotic linear behaviour $\varepsilon(\theta_0, P) \simeq +8 P \theta$, $P>0$, at $\theta \to - \infty$. Eventually, the $QQ$ system~\eqref{QQ0} characterizes the QNMs as $Y(\th_n-i\pi/2)=-1$, {\it i.e.} the {\it TBA quantization condition} 
\be \label{quanteps0}
\varepsilon(\th_{n'} -i \pi/2) = - i \pi(2n'+1) \,, \qquad n' \in \mathbb{Z}\,
\ee 
which can be easily implemented by using the TBA~\eqref{TBA0}  as table I shows. These values match very well with those obtained by the standard method of continued fractions by Leaver and is consistent with the ($l \to \infty$) WKB approximation (geodetic method) ~\cite{Leaver:1985,BianchiConsoliGrilloMorales:2021}.
\begin{table}[t]\label{QNMsD3brane}
\centering
\begin{tabular}{c|c|c|c|c}
$n$&$l $& TBA&Leaver& WKB \\
\hline
$0$&$0$&$\underline{1.369}12\, -\underline{0.504}048 i$&$\underline{1.369}72-\underline{0.504}311  i$&$1.41421\, -0.5 i$\\
$0$&$1 $&$\underline{2.091}18\, -\underline{0.501}788 i$&$\underline{2.091}76-\underline{0.501}811 i$&$2.12132\, -0.5 i$\\
$0$&$2$& $\underline{2.80}57\, -\underline{0.50100}9 i$&$\underline{2.80}629-\underline{0.50100}0 i$&$2.82843\, -0.5 i$\\
$0$&$3 $& $\underline{3.517}23\, -\underline{0.5006}49 i$&$\underline{3.517}83-\underline{0.5006}34 i$&$3.53553\, -0.5 i$\\
$0$&$4 $& $\underline{4.227}28\, -\underline{0.5004}53 i$&$\underline{4.227}90-\underline{0.5004}38 i$&$4.24264\, -0.5 i$
\end{tabular}
\caption{Comparison of QNMs of the D3 brane from TBA~\eqref{TBA0} (through~\eqref{quanteps0} with $n'=0$), Leaver (continued fractions) method and WKB (geodetic) approximation ($L=1$).}
\end{table}

We note that the physical condition $\Im \omega <0$ becomes by~\eqref{par0} $ - \pi/2 +2\pi n< \Im \theta < \pi/2+ 2 \pi n$, for $n \in \mathbb{Z}$. However, the TBA~\eqref{TBA0} is valid only for the fundamental strip $|\Im \th |< \pi/2$. In fact, in this region we find directly the QNMs for overtone number $n=0=n'$. We expect that analytically continuing the TBA by using the $Y$-system (\ref{Ysystem0}) in the other strips $|\Im (\th-2 \pi  i n) |< \pi/2$, we would obtain the other overtone numbers. We leave more details on this for future work.

Within our set-up of functional and integral equations for entire functions in $\theta$ (integrability), we can find other quantization conditions on the roots $\theta_n$ (QNMs). For instance, the $TQ$ relation~\cite{FioravantiGregori:2019}
\begin{equation}
T(\theta)Q(\theta) = Q(\theta - i \pi /2) +Q(\theta + i \pi /2) \, \label{TQ0}
\end{equation} 
means $Q(\theta_n - i \pi /2) +Q(\theta_n + i \pi /2) =0$. This and the $QQ$ relation (\ref{QQ0}) actually fixes $Q(\th_n + i \pi/2) Q(\th_n - i \pi/2 ) = 1$ and then 
\be \label{quantQ0}
Q(\th_n \pm i \pi/2) = \pm i 
\ee
are fixed, too. Again (\ref{QQ0}) around $\th_n$ forces $Q(\th + i \pi/2) =  i \pm   Q(\th)+\dots$ and $Q(\th - i \pi/2) =   - i \pm Q(\th)+\dots$ up to smaller corrections (dots). Therefore,~\eqref{TQ0} imposes 
\be \label{quantT0}
T(\th_n) =\pm 2\,.
\ee 
On equating the invariant trace of the monodromy matrix in the Floquet and ODE/IM bases (the coefficients in the latter involve the $T$ function), we obtain a nice equality involving the period $a$ ($=\nu$, Floquet index)~\cite{FioravantiGregori:2019}
\be  \label{Tnua}
T(\th) = 2 \cos 2 \pi a \, .
\ee 
In conclusion, condition~\eqref{QNMBethe} means that also the period $a$ is quantised 
\small
\be  \label{QuantA0}
a(\th_n) = \frac{n}{2} \,, \qquad n \in \mathbb{Z}\,\,.
\ee 
\normalsize
This is exactly the condition used by \cite{BianchiConsoliGrilloMorales:2021}. Yet, here we have fixed the general limits of its validity as relying on specific forms of the $TQ$ and $QQ$ systems \eqref{TQ0} and (\ref{QQ0}) respectively: it does not work in general, but we will see in the next section the specific conditions for its validity.

\section{IV. General theory: intersecting D3 branes and beyond}
\label{intD3branes}
To illustrate how our method develops in general, we consider the spacetime given by the intersection of four stacks of D3-branes in type IIB supergravity. These geometries are characterised by four different charges $\mathcal{Q}_i$ which, if all equal, lead to an extremal RN BH. In isotropic coordinates the line element writes
\small
\be  \label{lineintD3}
d s^2 = - f(r) d t^2 + f(r)^{-1}[d r^2 +r^2 (d \theta^2 + \sin^2 \theta d \phi^2)]\,,
\ee
with
$f(r) = \prod_{i=1}^4 \left(1 + \mathcal{Q}_i/r\right)^{-\frac{1}{2}}\,.$
\normalsize
The ODE describing the scalar perturbation is, with \small $\Sigma_k =\sum_{i_1 <...<i_k}^4 \mathcal{Q}_{i_1}\cdots \mathcal{Q}_{i_k}$~\cite{Ikeda:2021,BianchiConsoliGrilloMorales:2021b}
\small
\be  \label{ODEgrav2S}
\frac{d^2 \phi}{d r^2}+\left [ -\frac{(l+\frac{1}{2})^2-\frac{1}{4}}{r^2}+\omega^2\sum_{k=0}^4 \frac{\Sigma_k}{r^k}\right ]\phi = 0\,.
\ee 
\normalsize
\normalsize
Changing variables as 
$r=\sqrt[4]{\Sigma_4} e^{y} $ and 
\small
\be \label{DictIntD3branes}
\ba
 \omega \sqrt[4]{\Sigma_4} &= - i e^{\th} \,\,\,M_j = \frac{1}{2} \frac{\Sigma_{2j-1}}{\sqrt[4]{\Sigma_4}^{2j-1}} e^\th\,\,\,P^2=(l+\frac{1}{2})^2- \omega^2 \Sigma_2\,, 
\ea 
\ee 
\normalsize
($j=1,2$) the ODE takes the form 
\small 
\be  \label{ODEint11}
-\frac{d^2}{d y^2} \psi + \left[e^{2 \th}(e^{2y}+e^{-2y})+ 2e^{\th}( M_1 e^{y}+M_2 e^{-y}) +P^2 \right] \psi = 0\,.
\ee 
\normalsize
It generalizes that for the Perturbed Hairpin IM~\cite{FateevLukyanov:2005} and may have its own 2D statistical field theory interpetation. Connection with $SU(2)$ $N_f=2=(1,1)$ gauge theory is realized by flavour masses $m_j = \hslash M_j$, modulus $u= P^2/\hslash^2$ and instanton coupling $\L_2 = 4  \hslash e^\th$~\cite{ItoKannoOkubo:2017,BianchiConsoliGrilloMorales:2021b}.

Now, we develop the ODE/IM procedure for this general model by extending the porcedure of~\cite{FioravantiGregori:2019}. The regular solutions at $y \to \pm \infty$ are determined by
\small
\be  \label{asyreg2} 
\ba 
\psi_{\pm,0}(y) &\simeq 2^{-\frac{1}{2}-M_{\pm}}e^{-(\frac{1}{2}+M_{\pm})\th\mp(\frac{1}{2}+M_{\pm})y}e^{- e^{\th\pm y}}\,,\quad y\to \pm \infty\,, 
\ea 
\ee 
\normalsize
with $M_+=M_1$ and $M_-=M_2$. Equation~\eqref{ODEint11} enjoys the discrete symmetries
\small
\be 
\ba 
\Omega_{\pm} &: \,\, \th \to \th+i \pi/2 \,, \,\, y \to y\pm i \pi/2\,,\,\, M_1 \to \mp M_1\,,\,\, M_2 \to \pm M_2\,,  
\ea 
\ee 
\normalsize
which are consistent with the brane dictionary~\eqref{DictIntD3branes}. (The brane parameters vary as \small$\Sigma_1 \to \pm i \Sigma_1\,,\Sigma_2 \to - \Sigma_2\,,\Sigma_3 \to \mp i \Sigma_3\,, \Sigma_4 \to \Sigma_4$.) \normalsize
 Thanks to these symmetries , one can define other independent solutions 
$\psi_{-,k} = \Omega_{-}^k \psi_{-,0} \,,$ $\psi_{+,k}= \Omega_{+}^k \psi_{+,0}$. We also have the invariance properties $ \Omega_{+}^k \psi_{-,0}=\psi_{-,0} \,,$ $  \Omega_{+}^k \psi_{+,0}=\psi_{+,0}$.
These solutions are normalized such that their wronskians are
$W[\psi_{-,k+1},\psi_{-,k}]=  -i \exp \{ (-1)^k i \pi M_2 \}$, $W[\psi_{+,k+1},\psi_{+,k}]=  i \exp \{ (-1)^k i \pi M_1 \}$.
One can now define a Baxter's $Q$ function as the wronskian
\be 
 Q_{+,+}(\th,P) =W[\psi_{+,0},\psi_{-,0}](\th,P,M_1,M_2)\,.
\ee 
\small(We will use the notation $Q_{\pm,\pm} = Q(\th,\pm M_1,\pm M_2)$, $Q_{\pm,\mp} = Q(\th,\pm M_1,\mp M_2)$.)
\normalsize
By the properties of wronskians, we can write the linear connection relations as
\small
\be 
\ba  \label{conn2+-}
i e^{i \pi M_1}\psi_{-,0}&=Q_{-,+}(\th+i\pi/2)\psi_{+,0}- Q_{+,+}(\th)\psi_{+,1} \\
i e^{i \pi M_1}\psi_{-,1}&=Q_{-,-}(\th+i\pi)\psi_{+,0}-Q_{+,-}(\th+i\pi/2)\psi_{+,1}
\ea 
\ee
\normalsize
and taking their wronskian we obtain the $QQ$ system 
\small
\be  \label{QQ2}
\ba
Q_{+,-}(\th+\frac{i \pi}{2})Q_{-,+}(\th-\frac{i \pi}{2})
= e^{-i\pi (M_1-M_2)} +Q_{-,-}(\th)Q_{+,+}(\th)\,.
\ea
\ee 
\normalsize
One can define a $Y$ function as
\small $Y_{2,+,+}(\th) = \exp \{ i \pi( M_1 -M_2)\}Q_{+,+}(\th)Q_{-,-}(\th)\,$\normalsize and the $Y$ system is then
\small
\be 
\ba \label{Ysyst2}
Y_{+,-}(\th+\frac{i \pi}{2} )Y_{-,+}(\th-\frac{i \pi}{2} ) = [1 + Y_{+,+}(\th)][1 + Y_{-,-}(\th)]\,.
\ea
\ee 
\normalsize
In gravity variables it reads
\small
\be  \label{Ysyst2grav}
\ba
&Y(\th+ \frac{i\pi}{2},-i\Sigma_1,-\Sigma_2,i\Sigma_3)Y(\th-\frac{i\pi}{2},-i \Sigma_1,-\Sigma_2,i\Sigma_3 )\\
&= [1 + Y(\th,\Sigma_1,\Sigma_2,\Sigma_3 )][1 + Y(\th,-\Sigma_1,\Sigma_2,-\Sigma_3 )]
\ea
\ee 
\normalsize
and can be inverted into the TBA
\small
\be \label{TBA2}
\ba 
\varepsilon_{\pm,\pm}(\th)&= [f_{0,+}\mp \frac{i \pi}{2} (\frac{\Sigma_1}{\Sigma_4^{1/4}}-\frac{\Sigma_3}{\Sigma_4^{3/4}})]e^{\th}- \varphi \ast (\bar{L}_{\pm \pm}+ \bar{L}_{\mp \mp})(\th) \\
\bar{\varepsilon}_{\pm,\pm}(\th)&= [\bar{f}_{0,+}\pm \frac{\pi}{2}(\frac{\Sigma_1}{\Sigma_4^{1/4}}+\frac{\Sigma_3}{\Sigma_4^{3/4}})]e^{\th}- \varphi \ast (L_{\pm \pm}+ L_{\mp \mp})(\th) \\
\ea 
\ee 
\normalsize
where we defined
\small $\varepsilon(\th) =-\ln Y(\th,\Sigma_1,\Sigma_2, \Sigma_3,\Sigma_4)$, $\bar{\varepsilon}(\th) = \varepsilon(\th,i\Sigma_1,-\Sigma_2,-i \Sigma_3,\Sigma_4)$, $L = \ln [1+\exp \{ -\varepsilon \} ]$, $\varphi(\th) = (2\pi \cosh(\th))^{-1}$\normalsize. 
The forcing term is obtained from the relation $Q_{+,+}(\th)=- i e^{i \pi M_1} \lim_{y\to+\infty} \psi_{-,0}(y)/\psi_{+,1}(y)$ (following from~\eqref{conn2+-}) and we define $f_{0,\pm}=c_{0,+,\pm} +c_{0,-,\mp}$ and \small
$c_{0,\pm,\pm}= \int_{-\infty}^\infty\Bigl[\sqrt{2 \cosh(2 y) \pm\frac{\Sigma_1}{ \sqrt[4]{\Sigma_4}}e^y\pm\frac{\Sigma_3}{\sqrt[4]{\Sigma_4^3}}e^{-y}+\frac{\Sigma_2}{\sqrt{\Sigma_4}}} -\text{reg.}\Bigr] \, dy\,\, $
\normalsize
, which in turn can be expressed either through a triple power series for small parameters or as an elliptic integral.
We have to numerically input $l$ in the TBA with the boundary condition at \small $\th \to  -\infty$: $\varepsilon_{\pm,\pm}(\th) \simeq 4 P \th \simeq 4 (l+1/2) \th$\normalsize, also following from the asymptotic of the ODE~\eqref{ODEint11} (the precision improves by adding also the constant at the subleading order). This TBA is a generalization of those found in~\cite{FateevLukyanov:2005,Imaizumi:2021} for the Perturbed Hairpin IM and the $N_f=2$ gauge theory with equal masses $m_1 = m_2$ respectively, although we shall pay particular attention to the change of variables from gravity or gauge to integrability: this results in different TBA equations as first noted in~\cite{FioravantiGregori:2019}. Through this TBA we find again the QNMs to be given by the Bethe roots condition 
\be \label{quanteps2}
\bar{\varepsilon}_{+,+}(\th_{n'} -i \pi/2) = - i \pi (2n'+1) \,,\quad Q_{+,+}(\th_n) = 0
\ee
and we show in tables II and III their agreement with continued fraction (Leaver) method and WKB approximation ($l \to \infty$)~\cite{Leaver:1985,Ikeda:2021,BianchiRusso:2021}. For $\Sigma_1 \neq \Sigma_3$ and $\Sigma_4 \neq 1$ the Leaver method is not applicable, at least in its original version, as the recursion produced by the ODE involves more than $3$ terms ({\it cf.}~\cite{BianchiConsoliGrilloMorales:2021,Leaver:1985}) and thus also for this reason the TBA method may be regarded as convenient. However, there is a recent development, the so-called matrix Leaver method, which is still applicable~\cite{Leaver:1990,KumarBannonPribytokRodgers:2020}.
\begin{table}[t]
\centering
\begin{tabular}{c|c|c|c|c}
$n$&$ l $&  TBA&Leaver& WKB\\
\hline
$0$&1&$
 \underline{0.86}9623\, -\underline{0.372}022 i$&$\underline{0.86}8932-\underline{0.372}859 i$&$0.89642-0.36596 i$\\
$0$& 2&$\underline{1.477}990\, -\underline{0.368}144 i$&$\underline{1.477}888-\underline{0.368}240 i$&$1.4940-0.36596 i$\\
$0$&3&$\underline{2.080}200\, -\underline{0.3670}76 i$&$\underline{2.080}168-\underline{0.3670}97 i$&$2.0916-0.36596 i$\\
$0$&4&$\underline{2.6803}63\, -\underline{0.3666}37 i$&$\underline{2.6803}50-\underline{0.3666}42 i$&$2.6893-0.36596 i$\\
\end{tabular}
\caption{Comparison of QNMs obtained from TBA~\eqref{TBA2}, Leaver method (through~\eqref{quanteps2} with $n'=0$) and WKB approximation ($\Sigma_1 = \Sigma_3 =0.2 $, $\Sigma_2=0.4$, $\Sigma_4=1$.)}
\begin{tabular}{c|c|c|c|c}
$n$&$ l $&  TBA&Leaver& WKB\\
\hline
$0$&1&$
 0.896681\, -0.40069 i$& N.A. &$0.93069-0.39458 i$\\
$0$& 2&$1.5308\, -0.39676 i$&N.A. &$1.5511-0.39458 i$\\
$0$&3&$2.15708\, -0.395689 i$&N.A. &$2.1716-0.39458 i$\\
$0$&4&$2.78077\, -0.39525 i$&N.A. &$2.7921-0.39458 i$\\
\end{tabular}
\caption{Comparison of QNMs obtained from TBA~\eqref{TBA2},  (through~\eqref{quanteps2} with $n'=0$) and WKB approximation ($\Sigma_1 = 0.1$, $\Sigma_2 =0.2 $, $\Sigma_3=0.3$, $\Sigma_4 = 1$). Since $\Sigma_1 \neq \Sigma_3$ the Leaver method is not applicable (N.A.) in its original version.}\label{tabQNMsNf=2}
\end{table}

From ODE/IM we have evidence that generalizations of formula~\eqref{Q0GGM} can be derived (some are already available for the limit $N_f=1$~\cite{GrassiHaoNeitzke:2021}). In this way we expect that in general the integrability Bethe roots condition~\eqref{quanteps2}, which we have shown to follow straightforwardly from BHs physics, corresponds in gauge theory to the quantization of the $A_D$ period conjectured in~\cite{AminovGrassiHatsuda:2020}. 

The ODE/IM construction goes further because the presence of the irregular singularity of~\eqref{ODEint11} at $y \to \pm \infty$ (Stokes phenomenon) allow us to define $T$ functions  
\be 
T_{+,+}(\th) = - i W[\psi_{-,-1},\psi_{-,1}]\,, \,\, \tilde{T}_{+,+}(\th) =  i W[\psi_{+,-1},\psi_{+,1}]\,.
\ee 
Then, on expanding $\psi_{\pm,1}$ in terms of $\psi_{\pm,0}$, $\psi_{\pm,-1}$, similarly to~\eqref{conn2+-}, the $TQ$ relations follow
\small
\be 
\ba\label{TQNf2}
T_{+,+}(\th) Q_{+,+}(\th) &= e^{i \pi M_2} Q_{+,-}(\th-\frac{i \pi}{2})+e^{-i \pi M_2} Q_{+,-}(\th+\frac{i \pi}{2})\\
\tilde{T}_{+,+}(\th) Q_{+,+}(\th) &= e^{i \pi M_1} Q_{-,+}(\th-\frac{i \pi}{2})+e^{-i \pi M_1} Q_{-,+}(\th+\frac{i \pi}{2})\,.
\ea
\ee  
\normalsize
As above (\ref{Tnua}), we equate the invariant trace of the monodromy matrix in the Floquet and ODE/IM bases (for the coefficients in the latter the $T$s are needed) and obtain equalities involving now also the masses $M_1$ and $M_2$ 
\be \label{TnuNf=2}
\ba
2 \cos 2\pi \nu + 2 \cos 2 \pi M_1 &= \tilde{T}_{+,+}(\th )\tilde{T}_{-,+}(\th+i\frac{\pi}{2}) \\
2 \cos 2\pi \nu + 2 \cos 2 \pi M_2 &= T_{+,+}(\th)T_{+,-}(\th+i \frac{\pi}{2}) \,.
\ea
\ee 
This generalizes a simpler formula conjectured in ~\cite{FateevLukyanov:2005} and should give a proof method for the one guessed in ~\cite{FioravantiPoghossian:2019}. Besides, also in this case the period $a=\nu$, the Floquet index.

Only for {\it equal masses} $M_1=M_2\equiv M$ we can make considerations on $TQ$ and $QQ$ systems (\ref{TQNf2},\ref{QQ2}) like in section III and conclude that 
\be \label{TTqNf=2}
T_{+,+}(\th_n)T_{-,-}(\th_n) = 4 \, ,
\ee
which generalizes~\eqref{quantT0}. Then, we use $T_{+,-}(\th+i \frac{\pi}{2})= T_{+,+}(\th)$, $\tilde{T}_{-,+}(\th+i\frac{\pi}{2})= \tilde{T}_{+,+}(\th)$ and ~\eqref{TTqNf=2} in ~\eqref{TnuNf=2} (invariant upon changing sign of $M$) and conclude 
\be 
[ \cos 2\pi \nu + \cos 2 \pi M ]_{\theta=\theta_n}= \pm 2 \, .
\ee
Still for equal masses, there is the extra symmetry $\psi_{+,0}(y) = \psi_{-,0}(-y)$ as in the previous $N_f=0$ case. It exchanges infinity ($y=\infty$) and the (analogue) horizon ($y=- \infty$), leaving the photon-sphere ($y=0$) fixed as in~\cite{BianchiRusso:2021} and imposes the identity $\tilde{T}_{+,+}(\th) = T_{+,+}(\th)$.

\medskip
\section{V. Perspectives}

From the generality of the case of previous section it is manifest that our method applies to many other theories both to the angular and the radial problems~\cite{BianchiConsoliGrilloMorales:2021b}. We can list, for instance, the $SU(2)$ $N_f=(2,0)$ gauge theory and the associated gravity counterparts, like D1D5 fuzzballs and CCLP 5D BHs, as well as to $SU(2)$ $N_f=3$ gauge theory and the general asymptotically flat Kerr-Newman BH~\cite{BianchiConsoliGrilloMorales:2021b}. Instead of both irregular singularities as above, in these cases the ODE shows a regular and an irregular point, as in the original correspondence for conformal minimal models~\cite{DoreyTateo1998}: the associated integrability structure may share some similarity with the latter case. Another simple but tricky case happens when all the singularities are regular, e.g.  $SU(2)$ $N_f=4$ or asymptotically AdS BHs, since then the ODE has only regular singularities and no Stokes sectors. Actually, our procedure should still apply though trivialize as in \cite{DTS} for the supersymmetric XXZ spin chain. Yet, more recently these regular equations were discovered to appear (as they should) as ODEs in the spectral parameter, bi-spectral to another irregular ODE~\cite{FioravantiRossi:2021}: this bi-spectrality map is a route to be followed.

In any case, the TBA method seems in many cases convenient with respect to other methods, such as the original Leaver or SW methods, which cannot always be easily applied. The latter for instance needs a clever numerical re-summation procedure~\cite{AminovGrassiHatsuda:2020}.

Eventually, we note that much of the BH theory seems to go in parallel to the ODE/IM correspondence construction and its 2D statistical field theory interpretation, beyond the determination of QNMs: as an example, the {\it absorption coefficient} seems a ration of $Q$s. To this aim it is relevant to understand the r\^ole of excited states of ODE/IM ~\cite{BLZexc,Fexc} which preserve the form of the functional equations, but change the integral ones.

\medskip

\begin{acknowledgements}
\small {\bf Acknowledgements} We thank M. Bianchi, D. Consoli, A. Grillo, F. Morales for suggestions, and H. Shu, K. Zarembo for discussions. DG thanks NORDITA for warm hospitality.
\normalsize
\end{acknowledgements}



\begin{thebibliography}{xx}


\bibitem{LIGOScientific:2016}
LIGO Scientific and Virgo Collaborations - Abbott, B. P. and others,
 Phys. Rev. Lett. 116.061102 064 (2016), arXiv:1602.03837[gr-qc];
\bibitem{Cardoso:2017cqb}
Cardoso, V. and Pani, P., Nature Astron 1(2017)586, arXiv:1709.01525[gr-qc];
\bibitem{BianchiConsoliGrilloMorales:2021}
Bianchi, M. and Consoli, D. and Grillo, A. and Morales, Jos\`e F., arXiv:2105.04245[hep-th];
\bibitem{Mayerson:2020}
Mayerson, D. R., Gen. Rel. Grav. 52(2020)115, arXiv:2010.09736[hep-th];
\bibitem{AminovGrassiHatsuda:2020}
Aminov, G. and Grassi, A. and Hatsuda, Y., arXiv:2006.06111[hep-th];
\bibitem{SeibergWitten:1994b}
Seiberg, N. and Witten, E., Nucl. Phys. B431(1994)484, arXiv:9408099[hep-th];
\bibitem{NekrasovShatashvili:2009}
Nekrasov,N. and Shatashvili,S., ICMP09, 265, arXiv:0908.4052[hep-th]
\bibitem{BianchiConsoliGrilloMorales:2021b}
Bianchi, M. and Consoli, D. and Grillo, A. and Morales, Jos\`e F., arXiv:2109.09804[hep-th];
\bibitem{CasalsCosta:2021}
Casals, M. and da Costa, R. T., arXiv:2105.13329[gr-qc];
\bibitem{Hatsuda:2020Teukolsky}
Hatsuda, Y., arXiv:2007.07906[gr-qc];
\bibitem{BonelliIossaLichtigTanzini:2021}
Bonelli, G. and Iossa, C. and Lichtig, D. Panea and Tanzini, A., arXiv:2105.04483[hep-th];
\bibitem{AldayGaiottoTachikawa:2010}
Alday, L. F. and Gaiotto, D. and Tachikawa, Y. , Lett. Math. Phys. 91(2010)167, arXiv:0906.3219[hep-th];
\bibitem{FioravantiGregori:2019}
Fioravanti, D. and Gregori, D., Phys. Lett. B 804(2020)135376, arXiv:1908.08030[hep-th];
\bibitem{FioravantiPoghossian:2019}
Fioravanti, D. and Poghosyan, H. and Poghossian, R. , JHEP03(2020)049, arXiv:1909.11100[hep-th];
\bibitem{DoreyTateo1998}
Dorey, P. and Tateo, R., J. Phys. A32(1999)L419, arXiv:9812211[hep-th];
\bibitem{BazhanovLukyanovZamolodchikov2001}
Bazhanov, V. and Lukyanov, S. and Zamolodchikov, A. , J.Stat.Phys.102(1999)567, arXiv:9812247[hep-th];
\bibitem{DoreyTateo1999}
Dorey, P. and Tateo, R., Nucl.Phys. B563(1999)573, arXiv:9906219[hep-th];
\bibitem{FioravantiRossi:2021}
Fioravanti, D. and Rossi, M. arXiv:2106.07600[hep-th];
\bibitem{Ikeda:2021}
Ikeda, T. and Bianchi, M. and Consoli, D. and Grillo, A. and Morales, J. F. and Pani, P. and Raposo, G., Phys. Rev. D.104(2021)066021, arXiv:2103.10960[hep-th];
\bibitem{Nollert:1999}
Nollert, H.P., Class. Quant. Grav. 16(1999)R159;
\bibitem{GubserHashimoto:1998}
Gubser, S. S. and Hashimoto, A., Commun. Math. Phys. 203(1999)325, arXiv:9805140[hep-th];
\bibitem{ZamolodchikovMemorial}
Zamolodchikov, Al. B., \textit{Quantum Field Theories in two dimensions, 2}, World Scientific (2012); 
\bibitem{Bazhanov:1996dr}
V.~V.~Bazhanov, S.~L.~Lukyanov and A.~B.~Zamolodchikov,
Commun. Math. Phys.190(1997)247, arXiv:9604044[hep-th];
\bibitem{GrassiGuMarino}
Grassi, A. and Gu, J. and Marino, M. JHEP07(2020) 106, arXiv:1908.07065[hep-th];
\bibitem{Nekrasov:2002}
Nekrasov, N. A. , Adv. Theor. Math. Phys. 7(2003)831, arXiv:0206161[hep-th];
\bibitem{NekrasovOkounkov:2003}
Nekrasov, N. A. and Okounkov, A. , Prog. Math.  244(2006)525, arXiv:0306238[hep-th];
\bibitem{Matone:1995}
Matone, M., Phys. Lett. B357(1995)342, arXiv:9506102[hep-th];
\bibitem{FlumeFucitoMoralesPoghossian:2004}
Flume, R. and Fucito, F. and Morales, J. F. and Poghossian, R. , JHEP08(2004)004, arXiv:0403057[hep-th];
\bibitem{Leaver:1985}
Leaver, E. W., Proc.Roy.Soc.Lond. A402(1985)285;
\bibitem{FateevLukyanov:2005}
Fateev, Vladimir A. and Lukyanov, Sergei L., J. Phys. A 39(2006)12889, arXiv:0510271[hep-th];
\bibitem{ItoKannoOkubo:2017}
Ito, K. and Kanno, S. and Okubo, T, JHEP08(2017)065, arXiv:1705.09120[hep-th];
\bibitem{Imaizumi:2021}
Imaizumi, K. , Phys. Lett. B816(2021)136270, arXiv:2103.02248[hep-th];
\bibitem{Leaver:1990}
Leaver, E. W. , PhysRevD.41(1990)2986;
\bibitem{KumarBannonPribytokRodgers:2020}
Prem Kumar, S. and O'Bannon, A. and Pribytok, A. and Rodgers, R. JHEP05(2021)109, arXiv:2011.13859[hep-th];
\bibitem{GrassiHaoNeitzke:2021}
Grassi, A. and Hao, Q. and Neitzke, A. , arXiv:2105.03777[hep-th];
\bibitem{BianchiRusso:2021}
Bianchi, M., Di Russo, G. , arXiv:2110.09579[hep-th];
\bibitem{DTS}
Dorey, P. and Suzuki, J. and Tateo, R. , J.Phys.A37(2004)2047, arXiv:0308053[hep-th];
\bibitem{BLZexc}
Bazhanov, V. V. and Lukyanov, S. L. and Zamolodchikov, Al. B. ,  Adv. Theor. Math. Phys.4(2003)711, arXiv:0307108[hep-th];
\bibitem{Fexc}
Fioravanti, D., Phys. Lett. B609(2005)173, arXiv:0408079[hep-th].

\makeatletter
\providecommand \@ifxundefined [1]{%
 \@ifx{#1\undefined}
}%
\providecommand \@ifnum [1]{%
 \ifnum #1\expandafter \@firstoftwo
 \else \expandafter \@secondoftwo
 \fi
}%
\providecommand \@ifx [1]{%
 \ifx #1\expandafter \@firstoftwo
 \else \expandafter \@secondoftwo
 \fi
}%
\providecommand \natexlab [1]{#1}%
\providecommand \enquote  [1]{``#1''}%
\providecommand \bibnamefont  [1]{#1}%
\providecommand \bibfnamefont [1]{#1}%
\providecommand \citenamefont [1]{#1}%
\providecommand \href@noop [0]{\@secondoftwo}%
\providecommand \href [0]{\begingroup \@sanitize@url \@href}%
\providecommand \@href[1]{\@@startlink{#1}\@@href}%
\providecommand \@@href[1]{\endgroup#1\@@endlink}%
\providecommand \@sanitize@url [0]{\catcode `\\12\catcode `\$12\catcode
  `\&12\catcode `\#12\catcode `\^12\catcode `\_12\catcode `\%12\relax}%
\providecommand \@@startlink[1]{}%
\providecommand \@@endlink[0]{}%
\providecommand \url  [0]{\begingroup\@sanitize@url \@url }%
\providecommand \@url [1]{\endgroup\@href {#1}{\urlprefix }}%
\providecommand \urlprefix  [0]{URL }%
\providecommand \Eprint [0]{\href }%
\providecommand \doibase [0]{http://dx.doi.org/}%
\providecommand \selectlanguage [0]{\@gobble}%
\providecommand \bibinfo  [0]{\@secondoftwo}%
\providecommand \bibfield  [0]{\@secondoftwo}%
\providecommand \translation [1]{[#1]}%
\providecommand \BibitemOpen [0]{}%
\providecommand \bibitemStop [0]{}%
\providecommand \bibitemNoStop [0]{.\EOS\space}%
\providecommand \EOS [0]{\spacefactor3000\relax}%
\providecommand \BibitemShut  [1]{\csname bibitem#1\endcsname}%
\let\auto@bib@innerbib\@empty









\end{thebibliography}
\end{document}